\documentclass[a4paper,11pt]{article}
\usepackage{graphicx}
\usepackage{caption}
\usepackage{url}
\usepackage{xspace}
\newcommand{\accesstoken}{\textit{access\_token}\xspace}
\newcommand{\rarrow}{$\rightarrow$\xspace}
\newcommand{\idtoken}{\textit{id\_token}\xspace}

\newcommand{\authcodeflow}{\textit{Authorization Code Flow}\xspace}
\newcommand{\hybridserverflow}{\textit{Hybrid Server-side Flow}\xspace}
\newcommand{\clientflow}{\textit{Client-side Flow}\xspace}
\newcommand{\clientid}{\textit{client\_id}\xspace}
\newcommand{\clientsecret}{\textit{client\_secret}\xspace}
\begin{document}
\title{Analysing the Security of Google's implementation of OpenID Connect}
\author{Wanpeng Li and Chris J Mitchell\\
  Information Security Group,\\
  Royal Holloway,
  University of London\\
  TW20 0EX}
\date{\today}
\maketitle

\begin{abstract}
    Many millions of users routinely use their Google accounts to log in to relying party (RP) websites supporting the Google OpenID Connect service. OpenID Connect, a newly standardised single-sign-on protocol, builds an identity layer on top of the OAuth 2.0 protocol, which has itself been widely adopted to support identity management services. It adds identity management functionality to the OAuth 2.0 system and allows an RP to obtain assurances regarding the authenticity of an end user. A number of authors have analysed the security of the OAuth 2.0 protocol, but whether OpenID Connect is secure in practice remains an open question. We report on a large-scale practical study of Google's implementation of OpenID Connect, involving forensic examination of 103 RP websites which support its use for sign-in. Our study reveals serious vulnerabilities of a number of types, all of which allow an attacker to log in to an RP website as a victim user. Further examination suggests that these vulnerabilities are caused by a combination of Google's design of its OpenID Connect service and RP developers making design decisions which sacrifice security for simplicity of implementation. We also give practical recommendations for both RPs and OPs to help improve the security of real world OpenID Connect systems.

\end{abstract}
\section{Introduction} 
\label{sec:introduction}

In order to help alleviate the damage caused by identity-oriented attacks and simplify the management of identities, a range of identity management systems, such as OAuth 2.0, Shibboleth, CardSpace, OpenID, have been put forward \cite{cardspace, oauth2, openid}. As a replacement for the well-established OpenID  \cite{openid} scheme, OpenID Connect 1.0 \cite{openidconnect} builds an identity layer on top of the OAuth 2.0 framework \cite{oauth2}. The OAuth 2.0 framework enables an RP to obtain profile information about the end user, but does not provide any means for the RP to obtain information about the authentication of the end user. In OpenID Connect, in addition to obtaining profile information about the end-user, RPs can obtain assurances about the end user's identity from an OpenID Provider (OP), which itself authenticates the user.

OpenID Connect involves interactions between four core parties:

\begin{enumerate}
    \item the End User (U),  who attempts to access on-line services protected by the RP;
    \item the User Agent (UA), typically a web browser, that is employed by an end user to transmit requests to, and receive responses from, web servers;
    \item the OpenID Provider (OP), e.g.\ Google, which provides methods to authenticate an end user and generates assertions regarding the authentication event and the attributes of the end user;
    \item the Relying Party (RP), e.g.\ Wikihow, which provides protected on-line services and consumes the identity assertion generated by the IdP in order to decide whether or not to grant access to the end user.

\end{enumerate}

In summary, the end user employs a user agent to access resources provided and protected by the RP, which relies on the OP to provide authentic information about the user.

Even though OpenID Connect was only finalised at the start of 2014, there are already more than half a billion OpenID Connect-based user accounts provided by Google \cite{googleopenidconnect2015}, PayPal \cite{paypalopenidconnect2014} and Microsoft \cite{microsoftopenidconnect2014}. This large user base has led very large numbers of RPs to integrate their services with OpenID Connect.

The security of OAuth 2.0, the foundation for OpenID Connect, has been analysed using formal methods \cite{oauth2threat,pai11,frostig11}. Research focusing on implementations of OAuth 2.0 has also been conducted \cite{DBLP:conf/ccs/ChenPCTKT14, DBLP:conf/isw/LiM14, DBLP:conf/ccs/SunB12,DBLP:conf/sp/WangCW12,DBLP:conf/uss/ZhouE14}. However, as a newly standardised protocol, it is not yet clear whether practical implementations of OpenID Connect properly follow the specification \cite{openidconnect}.  Given the large scale use of the Google service, it is important to understand how secure deployments of OpenID Connect really are. In order to help answer the question, the operation of all one thousand sites from the GTMetrix Top 1000 Sites \cite{GTmetrix15} providing services in English was examined. Of these sites, 103 were found to support the use of the Google's OpenID Connect service, at the time of our survey (early 2015). All 103 of these websites were then further examined for potential vulnerabilities, with the results as reported in this paper. In our study, all the RPs and the Google OP site were treated as black boxes, and the HTTP messages transmitted between the RP and OP via the browser were carefully analysed to identify possible vulnerabilities. For every identified vulnerability, we implemented and tested an exploit to evaluate the possible attack surface.

Our study reveals serious vulnerabilities of a number of types, which either allow an attacker to log in to the RP website as the victim user or enable the compromise of potentially sensitive user information. Google has customised its implementation of OpenID Connect by combining SDKs, web APIs and sample code, and as a result the OpenID Connect specification only acts as a loose guideline to what RPs have actually implemented. Further examination suggests that the identified vulnerabilities are mainly caused by RP developers misunderstanding how to use the Google OpenID Connect service, and by making design decisions which sacrifice security for simplicity of implementation. Some of the attacks we have discovered use cross-site scripting (XSS) \cite{DBLP:conf/ndss/NadjiSS09,DBLP:conf/ndss/VogtNJKKV07,DBLP:conf/icse/WassermannS08,DBLP:conf/sac/KirdaKVJ06} and cross site request forgeries (CSRFs) \cite{DBLP:conf/ccs/BarthJM08,DBLP:conf/esorics/RyckDJP11,jovanovic2006preventing,DBLP:conf/fc/MaoLM09,zeller2008cross}, well-established
and widely exploited attack techniques.

OpenID Connect  is being used to protect millions of user accounts, as well as sensitive user information stored at both RPs and the Google OP server.  Moreover, as of April 20th 2015, Google shut down its OpenID 2.0 \cite{googleopenid2015} service; as a result a huge number of RPs have had to upgrade their Google sign-in service to use OpenID Connect. It is therefore vitally important that the issues we have identified are addressed urgently, and that Google considers issuing updated advice to all RPs using its service. In this connection we have notified all the RPs in whose OpenID Connect service we have identified serious vulnerabilities, as well as Google itself.

The remainder of the paper is organised as follows. In Section 2 we give an overview of OpenID Connect. We describe our adversary model in section 3. Section 4 describes the experiments we performed to evaluate the security of Google's implementation of OpenID Connect. Possible reasons for the identified vulnerabilities are discussed in section 5. In section 6 we give proposed mitigations for these vulnerabilities, we review related work in section 7, and section 8 concludes the paper.
\section{OpenID Connect} 
\label{sec:openid_connect}
\subsection{Introduction} 
\label{sub:introduction}

OAuth 2.0 \cite{oauth2} enables an end user to grant an online service controlled access to his or her personal details (e.g.\ email address, birth date, pictures) held by a third party, without revealing his or her password for this third party. Without the help of a system like OAuth 2.0, relying parties (e.g.\ Wikihow) would need to ask the user to provide his or her password in order to access user information stored at a third party (e.g.\ Google), which is a less than ideal solution.

As already noted, OpenID Connect 1.0 \cite{openidconnect} is built as an identity layer on top of the OAuth 2.0 protocol. The functionality that it adds enables RPs to verify the identity of an end user by relying on an authentication process performed by an OpenID Provider (OP), i.e.\ it adds identity management functionality to the OAuth 2.0 system.


\subsection{OpenID Connect Tokens} 
\label{sub:tokens_used_in_openid_connect}

In order to enable an RP to verify the identity of an end user, OpenID Connect adds a new type of token to OAuth 2.0, namely the \idtoken. This complements the \accesstoken and \textit{code}, which are already part of OAuth 2.0. These three types of token are all issued by an OP, and have the following functions.
\begin{itemize}
    \item A \textit{code} is an opaque value which is bound to an identifier and a URL of the RP\@. Its main purpose in OpenID Connect is as a means of giving an RP authorisation to retrieve tokens from the OP\@. In order to help minimise threats arising from its possible exposure, it has a limited validity period and is typically set to expire shortly after issue to the RP \cite{oauth2}.
    \item An \accesstoken contains credentials used to authorise access to protected resources stored at a third party (e.g.\ the OP). Its value is an opaque string representing an authorization issued to the RP\@. It encodes the right for the RP to access data held  by a specified thrid party with a specific scope and duration, granted by the end user and enforced by the RP and the OP\@.
    \item An \idtoken contains claims about the authentication of an end user by an OP together with any other claims requested by the RP\@. Claims that can be inserted into such a token include: the identity of the OP that issued it, the user's unique identifier at this OP, the identity of the intended recipient, the time at which it was issued, and its expiry time. It takes the form of a JSON Web Token \cite{jsonwebtoken2014} and is digitally signed by the OP\@.
\end{itemize}

Both an \accesstoken \cite{oauth2_client_side_google2015} and an \idtoken \cite{verify_id_token} can be verified by making a call to the web API of the issuing OP.
\subsection{Authentication Flows} 
\label{sub:authentication_flows}

\begin{figure}[htbp]
    \centering
    \includegraphics[width=\textwidth]{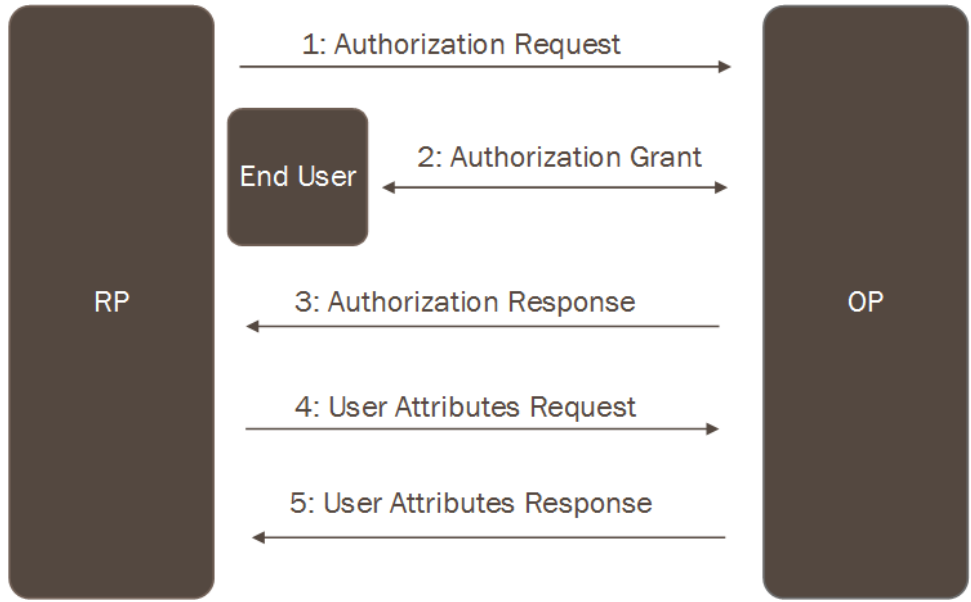}
    \caption{OpenID Connect Protocol Overview}
    \label{fig:openidconnect}
\end{figure}

OpenID Connect builds on user agent HTTP redirections. We suppose that an end user wishes to access services protected by the RP, which consumes tokens generated by the OP\@. The RP generates an authorization request on behalf of the end user and sends it to the OP via the UA, which is typically a web browser. The OP provides ways to authenticate the end user, asks the end user to grant permission for the RP to access the user attributes, and generates an authorization response which includes tokens of two types: \textit{access\_tokens} and \textit{id\_tokens}, where the latter contain claims about a user authentication event. After receiving an \accesstoken, the RP can use it to access end user's attributes using the API provided by the OP, and after receiving an \idtoken the RP is informed about the authentication of the user, as summarised in Fig. \ref{fig:openidconnect}.

Google's implementation of OpenID Connect \cite{openidconnect} supports four types of authentication flow \cite{googleopenidconnect2015}, i.e.\ ways in which the system can operate, namely \textit{Authorization Code Flow}, \textit{Hybrid Server-side Flow} (also knows as \textit{Hybrid Flow}) \cite{connect_server_side_google15}, \textit{Client-side Flow} (also known as \textit{Implicit Flow}), and \textit{Pure Server-side Flow}. However, as the \textit{Pure Server-side Flow} is rarely used and Google states that this flow is not recommended, we only give detailed descriptions of the first three flows.

The RP must register with the OP before it can use Google OpenID Connect. During registration, the OP gathers security-critical information about the RP, including the RP's redirect URI or \textit{origin}. The redirect URI is used in the \authcodeflow, and is the URI to which the user agent is redirected after step 5 (see step 5 of section \ref{subsubsec:authorization-code-flow}). The \textit{origin} is used in the \hybridserverflow and \clientflow, and is a pointer to the domain name of the RP. The OP issues the RP with a unique identifier (\textit{client\_id}) and a secret (\textit{client\_secret}) which it uses to authenticate the RP when using the \authcodeflow or \hybridserverflow.

\subsubsection{Hybrid Server-side Flow}
\label{subsubsec:hybrid-server-side-flow}

Google's OpenID Connect implementation uses \textbf{postMessage} \cite{w3cpostmessage,OpenIDConnectSessionManagement2014,DBLP:journals/cacm/BarthJM09,DBLP:conf/ndss/SonS13} to enable cross domain communication between the RP and Google's OP\@. Normally, scripts on different pages are only allowed to access each other if the web pages that caused them to execute are at locations sharing the same protocol, port number and host. The \textbf{postMessage} method lifts this restriction by providing a way to securely pass messages across domains. Further information on the operation of \textbf{postMessage} can be found, for example, in Son and Shmatikov \cite{DBLP:conf/ndss/SonS13}. In the \hybridserverflow and \clientflow, an RP Client is running on the UA and listening for the postMessage event.

We now give a detailed description of the \hybridserverflow.

\begin{enumerate}
    \item U \rarrow UA \rarrow RP: The user clicks the Google Sign-In button rendered on the RP website, and this causes the UA to send an HTTP or HTTPS request to the RP\@.
    \item \label{HybridSeverStep:request} RP \rarrow UA \rarrow OP: The RP generates an OpenID Connect authorization request and sends it to the OP via the UA. The authorization request includes  \textit{client\_id},  an identifier the RP registered with the OP previously;  \textit{response\_type=code token id\_token} which requests that a \textit{code}, an \accesstoken and an \idtoken be returned directly from Google; \textit{redirect\_uri=postmessage}, indicating that \textbf{postMessage} is being used; \textit{state}, an opaque value used by the RP Client to maintain state between the request and the callback (step \ref{HybridSeverStep:response} below); \textit{origin}, a URL without a path appended; and \textit{scope}, the scope of the requested permission.
    \item OP \rarrow UA: If the user has already been authenticated by the OP then this step and the next are skipped. If not, the OP returns a login form which is used to collect user authentication information (e.g.\ user account and password).
    \item U \rarrow UA \rarrow OP: The user completes the login form and grants permission for the RP to access the attributes stored by the OP.
    \item \label{HybridSeverStep:response} OP \rarrow UA: After receiving the permission grant from the user, the OP generates an HTML document which contains the authorization response and sends it back to the UA. The authorization response contains the \textit{code}, \textit{access\_token} and \textit{id\_token} generated by the OP; and \textit{state}, the value sent in step \ref{HybridSeverStep:request}.
    \item UA \rarrow RP: The UA executes the JavaScript inside the HTML document it received in the previous step. The JavaScript sends the authorization response using \textbf{postMessage} to the RP Client which is running on the UA and listening for the \textbf{postMessage} event. After the RP Client receives the authorization response it extracts the \textit{code} and sends it back to the RP.
    \item RP $\rightarrow$ OP: The RP produces an \accesstoken request and sends it to the OP token endpoint directly (i.e.\ not via the UA). The request includes \textit{grant\_type=authorization\_code}, indicating that the RP wants to use the \textit{code} to retrieve an \accesstoken from the OP; the \textit{code} generated in step \ref{AuthorizationCodeFlow:response}; \textit{redirect\_uri=postmessage}, indicating that \textbf{postMessage} has been used to get the \textit{code}; and \textit{client\_secret}, the secret shared by the RP and OP.
    \item OP $\rightarrow$ RP: The OP checks the \textit{code}, \textit{client\_secret} and \textit{redirect\_uri} and, if correct, responds to the RP with \textit{access\_token} and \idtoken, the latter of which is the same as the \idtoken sent in step \ref{HybridSeverStep:response}.
    \item RP $\rightarrow$ OP: The RP verifies the \idtoken. If it is valid, the RP now has evidence that the user has been authenticated. If necessary it can also make a web API call to retrieve the user attributes from the OP using the \accesstoken as evidence of its right to do so.
\end{enumerate}

In Google's implementation of \hybridserverflow, a \textit{code}, an \accesstoken and an \idtoken are always returned by Google to the RP's JavaScript client running on the user's browser. This means that these tokens are potentially revealed to the user agent and any applications which might be able to access the user agent.

\subsubsection{Client-side Flow}
The authorization request and response employed in the \clientflow are similar to those used in the \hybridserverflow. The only difference between the two flows is that in the \clientflow no \textit{code} is submitted back to the RP\@. The \clientflow operates as follows, where steps 1-5 are the same as steps 1-5 in section \ref{subsubsec:hybrid-server-side-flow}.
\begin{enumerate}
    \item[6.] UA \rarrow RP: The UA executes the JavaScript inside the html document it received in the previous step. The JavaScript sends the authorization response using \textbf{postMessage} to the RP Client that is running on the UA and listening for the \textbf{postMessage} event. After receiving the authorization response, the RP Client extracts the \accesstoken and \idtoken. It then verifies the \idtoken; if the \idtoken is valid, the RP now has evidence that the user has been authenticated. If necessary it can also make a web API call to retrieve the user attributes from the OP, using the \accesstoken as evidence of its right to do so.
    \item[7.] UA \rarrow U: The RP Client running on the UA updates the displayed web page based on the attributes it retrieved in the previous step.
\end{enumerate}

\subsubsection{Authorization Code Flow}
\label{subsubsec:authorization-code-flow}

One advantage of this flow is that no tokens are made available to the user agent or to any malicious applications which might be able to access the user agent. This is advantageous since if either of the tokens are obtained by a malicious party they could be used to access sensitive user data and/or successfully masquerade as the user. The OP must authenticate the RP before it issues the pair of tokens, and hence use of the \authcodeflow requires that an RP shares a secret with the OP\@.
The flow involves the OP returning an authorization \textit{code}, typically a short-lived opaque string, to the RP, which uses it to obtain the \idtoken and \accesstoken directly from the OP's \accesstoken endpoint, i.e.\ not via the UA\@. The main steps are as follows.

\begin{enumerate}
    \item U $\rightarrow $ RP: The user clicks a login button on the RP website, as displayed by the UA, which causes the UA to send an HTTP or HTTPS request to the RP.
    \item \label{AuthorizationCodeFlow:request} RP $\rightarrow$ UA $\rightarrow$ OP: The RP generates an OpenID Connect authorization request and sends it to the OP via the UA. The authorization request includes \textit{client\_id},  a client identifier which the RP registered with the OP previously;  \textit{response\_type=code}, indicating that the \authcodeflow is being used; \textit{redirect\_uri}, the URI to which the OP will redirect the UA after access has been granted; \textit{state}, an opaque value used by the RP to maintain state between the request and the callback (step \ref{AuthorizationCodeFlow:response} below); and \textit{scope}, the scope of the requested permission.
    \item OP $\rightarrow$ UA: If the user has already been authenticated by the OP then this step and the next are skipped. If not, the OP returns a login form which is used to collect user authentication information.
    \item U $\rightarrow$ UA $\rightarrow$ OP: The user completes the login form and grants permission for the RP to access the attributes stored by the OP.
    \item \label{AuthorizationCodeFlow:response}OP $\rightarrow$ UA: After using the information provided in the login form to authenticate the user, the OP generates an authorization response and sends it back to the UA\@. The authorization response contains \textit{code}, the authorization code generated by the OP; and \textit{state}, the value sent in step \ref{AuthorizationCodeFlow:request}.
    \item UA $\rightarrow$ RP: The UA redirects the response received in Step 5 to the RP.
    \item RP $\rightarrow$ OP: The RP produces an \accesstoken request and sends it to the OP token endpoint directly (i.e.\ not via the UA). The request includes  \textit{grant\_type=code}, indicating the RP wants to use the \textit{code} to retrieve an \accesstoken ; the \textit{code} sent in step \ref{AuthorizationCodeFlow:response}; the \textit{redirect\_uri}; and \textit{client\_secret}, the secret shared by the RP and OP.
    \item OP $\rightarrow$ RP: The OP checks the \textit{code}, \textit{client\_secret} and \textit{redirect\_uri} and if all are correct responds to the RP with an \textit{access\_token} and \textit{id\_token}.
    \item RP $\rightarrow$ OP: The RP verifies the \idtoken. If it is valid, the RP now has evidence that the user has been authenticated. If necessary it can also make a call to the web API offered by the OP, using the \accesstoken for authorisation, in order to retrieve the desired user attributes.
\end{enumerate}


\section{Adversary Model} 
\label{sec:adversary_model_and_related_work}

In our assessment of the security of Google's implementation of OpenID Connect, and of the implementations of specific RPs using the service, we consider two possible scenarios for the capabilities of an adversary.

\begin{itemize}
  \item \textbf{A Web Attacker} can share malicious links and/or post comments which contain malicious content (e.g.\ stylesheets or images) on a benign website; and/or exploit vulnerabilities in an RP website. The malicious content forged by a web attack might trigger the web browser to send HTTP/HTTPS requests to an RP and OP using either the GET or POST methods, or execute JavaScript scripts crafted by the attacker. For example, a web attacker could operate an RP website in order to try to collect \textit{access\_tokens}.

  \item \textbf{A Passive Network Attacker} has the ability to intercept unencrypted data sent between an RP and an end user browser (e.g.\ by monitoring an open Wi-Fi network).

\end{itemize}

Conducting a security analysis of commercially deployed OpenID Connect SSO systems requires a number of challenges to be addressed. These include lack of access to detailed specifications for the SSO systems, undocumented RP and OP source code, and the complexity of APIs and/or SDK libraries in deployed SSO systems. The methodology we used is similar to that employed by Wang et al.\ \cite{DBLP:conf/sp/WangCW12} and Sun and Beznosov \cite{DBLP:conf/ccs/SunB12}, i.e.\ we analysed the browser relayed messages (BRMs). We treated the RPs and IdPs as black boxes, and analysed the BRMs produced during authorization to look for possible exploit points. Since we used a black-box approach, there may very well be vulnerabilities, implementation flaws and attack vectors which our study did not uncover.


\section{A Security Study} 
\label{sec:studying_the_security_of_openid_connect}

To evaluate the security of OpenID Connect, we used Fiddler\footnote{\url{http://www.telerik.com/fiddler}} to capture BRMs sent between RPs and the OP;
we also developed a Python program to parse the BRMs to simplify analysis and to avoid mistakes resulting from manual inspections. All the experiments were performed using accounts set up specially for the purpose; i.e.\ at no time was any user's account accessed without permission. Of the 103 RPs supporting Google OpenID Connect that we examined, we found that 69 (67\%) adopt the \authcodeflow, 33 (32\%) use the \hybridserverflow, and just 1 adopted the \clientflow.

\subsection{Studying the security of the Hybrid Server-side Flow} 
\label{sub:exploiting_the_hybrid_server_side_flow}

As described in section \ref{subsubsec:hybrid-server-side-flow}, Google's OpenID Connect API uses \textbf{postMessage} to deliver the authorization response from the OP to an RP\@. When the RP Client running on the user's browser receives the authorization response from the OP, it extracts the  \textit{code} from the authorization response and then submits the \textit{code} back to the RP's OpenID Connect sign-in endpoint.

\subsubsection{Authentication by Google ID} 
\label{ssub:log_in_to_the_rp_as_the_victim_user}

As stated above, the RP's JavaScript client running on the UA submits the \textit{code} it receives from the Google OP back to the RP's Google sign-in endpoint (see step 6 in section \ref{subsubsec:hybrid-server-side-flow}). The \textit{code} plays a critical role in guaranteeing the user identity to the RP, in that the RP is meant to use it to retrieve the \accesstoken and \idtoken from the Google OP\@. However, we observed that 18\% of the RPs using the Hybrid Server-side Flow (i.e.\ 6 of the 33) submit the user's Google ID to the RP's Google sign-in endpoint; of these six RPs, two simply submit the user's Google ID without appending a \textit{code}, and one submits the user's Google ID with an \accesstoken. This led us to suspect that such RPs might be basing their verification of user identity solely on the Google ID, and not using the \textit{code} as it is intended to be used. If this were to be the case, then a web attacker which knows a user's Google ID could use it to log in to the user's RP account. We tested this, and found that as many as 9\% of the RPs using the Hybrid Server-side Flow (i.e.\ 3 of the 33) have this vulnerability.

It would appear that learning the Google ID for a victim user
can be relatively simple, as a user's Google+ post
URL reveals the user's Google ID\@.
An attacker can use the Google+ \textit{search for people} function
to find a victim user to attack, and can then visit the
chosen victim user's Google+ page to learn the ID\@. For example, \sloppy \url{https://plus.google.com/u/0/115722834054889887046/posts}
is the Google+ post URL for a Gmail account, for which
the ID is \url{115722834054889887046}.

We reported our findings to the three affected websites, and recommendations were also provided to enable the RP developers to fix the problem (see also \ref{sub:notifying_affected_parties}).


\subsubsection{Using the Wrong Token} 
\label{ssub:impersonation_the_victim_user}

An \accesstoken is a bearer token; this means that any party in possession of an \accesstoken can use it to get access to the associated user attributes stored by Google. This is the intended use of an \accesstoken; by contrast, the \idtoken is designed for use for providing assurances about user authentication. However, in practice, some RPs use an \accesstoken as a means of obtaining assurances about user authentication without verifying it (i.e.\ making a web API call to the OP token information endpoint \cite{oauth2_client_side_google2015}). In such a case, any party (e.g.\ another RP) that has obtained a user's \accesstoken can impersonate that user to the RP simply by submitting it. This is a particular threat in the case of a malicious RP, which can routinely obtain \textit{access\_tokens} from the Google OP\@. In other words, any RP using Google OpenID Connect has the ability to log in as a victim user to any RPs which use an \accesstoken to authenticate the user without verifying it. Unfortunately, we found that 58\% of RPs using the \hybridserverflow (i.e.\ 19 out of 33) submit an \accesstoken back to their Google sign-in endpoint (see step 6 in section \ref{subsubsec:hybrid-server-side-flow}) and 45\% (i.e.\ 15 out of these 19) use the \accesstoken to authenticate the user; of these 15 RPs, only two RPs verify the \accesstoken before using it to retrieve user attributes. As a result, 39\% of the RPs (i.e.\ 13 out of 33) that we examined are vulnerable to this impersonation attack.

We tested the above attack using Burp Suite\footnote{\url{http://portswigger.net/burp/}} by submitting an \accesstoken obtained from the 9GAG\footnote{{\url{http://9gag.com}}} website to the target RP's Google sign-in endpoint. If the attack succeeds, we are able to log in to the target RP as the victim user. As noted above, as many as 39\% of the RPs using the \hybridserverflow are vulnerable to this attack. Some of the vulnerable RPs (i.e.\ 3 out of 13) require additional evidence of the user to be submitted with the \accesstoken, in the form of the Google ID or the user's email address. However, an attacker that has an \accesstoken can very easily use it to obtain the user's Google ID and/or email address from Google, and so such additional steps do not prevent the impersonation attack.


\subsubsection{Intercepting an \accesstoken} 
\label{ssub:sniff_access_token}
As stated above, 58\% of RPs using the \hybridserverflow require the submission of an \accesstoken back to their Google sign-in endpoint (see step 6 in section \ref{subsubsec:hybrid-server-side-flow}). If the RP Client running on the UA sends an \textit{\accesstoken} back to its Google sign-in endpoint without SSL protection, a passive network attacker is able to intercept it (see section \ref{sec:adversary_model_and_related_work}). According to the OAuth 2.0 specification \cite{oauth2:bearer}, an \accesstoken should never be sent unencrypted between the user browser and the RP\@. However, we found that 12\% of RPs using the \hybridserverflow (i.e.\ 4 out of 33) send the \accesstoken unprotected. A sniffer written in Python was implemented to test this.

We also observed that one additional site, namely TheFreeDictionary\footnote{\url{http://www.thefreedictionary.com}} does use SSL to protect the transfer of the \textit{code} to its Google sign-in endpoint. However, the \accesstoken is subsequently stored in a cookie, and when the cookie is sent from the browser back to TheFreeDictionary the link is not SSL-protected. That is, the \accesstoken is observable by a passive eavesdropper.


\subsubsection{Privacy Issues} 
\label{ssub:privacy_issues}

When a user chooses to use OpenID Connect to log in to an RP website, the user attributes (e.g.\ email address, name) that the RP retrieves from the OP should never be revealed to parties other than the RP\@. SSL connections should be established to protect user information transmitted between the browser and the RP or OP\@.

However, as explored in greater detail below, user information leakage might happen if:
\begin{itemize}
    \item the RP Client running on the user's browser sends user information, the \idtoken or the \accesstoken back to its Google sign-in endpoint without SSL protection (see step 6 in section \ref{subsubsec:hybrid-server-side-flow});
    \item the RP Google sign-in endpoint sends the user information directly to the user's browser without SSL protection; or
    \item the RP uses SSL to protect the link to the Google sign-in endpoint, but changes to http when it sends the user's information back to the user's browser.
\end{itemize}

As described in Section \ref{ssub:sniff_access_token}, a passive eavesdropper can intercept the \accesstoken for 12\% of the RPs that use the Hybrid Server-side Flow (i.e.\ 4 out of 33), and can then use it to retrieve potentially sensitive user information, notably including the Google ID and email address. As stated in section \ref{sub:tokens_used_in_openid_connect}, the \idtoken is a JSON web token, in which the user email address and Google ID are encoded in cleartext using Base64; as a result anyone obtaining the token can immediately obtain the information within it. One of the four RPs referred to above sends an \idtoken in addition to the \accesstoken to its Google sign-in endpoint, and thus a passive web attacker can retrieve the user information it contains without requesting it from Google using the \accesstoken. We also found that one RP did not enable SSL to protect its Google sign-in endpoint and returned user information directly to the user browser. Another RP sends user information back to its Google sign-in endpoint without SSL protection. Yet another RP uses SSL to protect the link to the Google sign-in endpoint, but changes to HTTP when it sends the user information back to the user browser. As a result, user privacy cannot be guaranteed for  21\% of the RPs we examined (i.e.\ 7 out of 33). As noted above, a sniffer in Python was implemented to demonstrate the feasibility of the attack.

\subsubsection{Session Swapping} 
\label{ssub:swap_session}

As discussed earlier, the RP Client running on the UA sends the user's OpenID tokens (i.e.\ one or more of a \textit{code}, an \accesstoken, an \idtoken , and the user's Google ID) back to its Google sign-in endpoint (see step 6 in section \ref{subsubsec:hybrid-server-side-flow}). The OpenID Specification \cite{openidconnect} recommends that a \textit{state} value should be appended when the RP Client sends the tokens back to its Google sign-in endpoint, and that this \textit{state} value should be bound to the browser session.
If the RP Client fails to send the \textit{state} value, an attacker can execute a session swapping attack \cite{DBLP:conf/ccs/BarthJM08, DBLP:conf/ccs/SunB12,DBLP:conf/idman/DelftO10} by performing the following steps.
\begin{enumerate}
    \item The attacker first logs in to the RP website using his or her own account (see step 4 in section \ref{subsubsec:hybrid-server-side-flow}), and intercepts the tokens generated by Google (see step 5 in section \ref{subsubsec:hybrid-server-side-flow}) .
    \item The attacker constructs a request to the RP's Google sign-in endpoint including the attacker's own tokens.
    \item The attacker inserts the request in an HTML document (e.g.\ in the \textbf{src} attribute of a \textbf{img} or \textbf{iframe} tag) which is made publicly available via an HTTP server.
    \item The victim user is now, by some means, induced to visit the website offering the attacker's web page. The HTML can be constructed in such a way (described in detail below) that the victim's UA will automatically use the GET or POST method to send the attacker-constructed request to the RP; as a result the user session on the RP website will be bound to the attacker's account.
\end{enumerate}

We observed that 42\% of the RP Clients using the \hybridserverflow (i.e.\ 14 out of 33) use the POST method to submit the tokens back to the RP's server without an accompanying \textit{state} value. Use of a static \textbf{img} or \textbf{iframe} tag to perform an attack of the above
type does not work against these RPs, as the browser will automatically use the GET method to retrieve the img and iframe data. Thus, in order to use the POST method to submit those tokens, we created a special HTML page to conduct our session-swapping attack\@. We used JavaScript to create an iframe with a unique name in the browser. We then constructed a form inside the iframe whose action points to the RP's Google sign-in endpoint. We then put the attacker's tokens into the form input and configured the HTML to submit the form whenever the HTML document is loaded into a browser.

To deploy the attack, the constructed HTML page is made available via a publicly available web server. If a victim user visits this page, the JavaScript inside the HTML automatically submits the attacker's tokens to the RP using the POST method; as a result the victim user's session on the RP is bound to the attacker's, i.e.\ a session-swapping attack has been performed. An attacker could use such an attack to collect sensitive user information, e.g.\ if the victim user updates his credit card information on the RP website, the credit card information will be written to the attacker's account.

Unfortunately, we found that 73\% of RPs which adopt the \hybridserverflow (i.e.\ 24 out of 33) are vulnerable to this attack. Of these 24 RPs, eight (i.e.\ 24\% of this category) submit a \textit{code} to their Google sign-in endpoint; as the \textit{code} is a one-time value, the attacker must update it within the attack HTML every time the page is retrieved by a victim user.
For the other 48\% of vulnerable RPs (i.e.\ 16 out of 33), an \accesstoken or the user's Google ID is submitted back to the Google sign-in endpoint, in
which case the attacker does not need to update the attack page HTML as frequently.


\subsection{Studying the security of the Authorization Code Flow} 
\label{sub:studying_the_security_of_the_authorization_code_flow}

We start by observing that Google's OAuth 2.0 \textit{Authorization Code Flow} implementation \cite{google_oauth2_codeflow}  has similar steps to those given in \ref{subsubsec:authorization-code-flow}. The token endpoint provided as part of Google's implementation of OAuth 2.0 (as checked on April 22, 2015) returns an \textit{id\_token} to the RP\@. That is, without knowing details of the RP's internal operation, we cannot distinguish whether an RP is using OpenID Connect or OAuth 2.0. In this paper we therefore cover all cases where Google returns a \textit{code} to the RP's Google sign-in endpoint under our discussion of the OpenID Connect \authcodeflow, even though some of the RPs concerned may actually be using OAuth 2.0. However, this makes no difference to our security analysis.

Around 67\% of the RPs we examined (i.e.\ 69 out of 103) use the \authcodeflow. Unlike the \hybridserverflow, Google's implementation of \authcodeflow uses HTTP status code redirect techniques (i.e.\ using code 302) to deliver the authorization response to the RP's Google sign-in endpoint.

\subsubsection{Intercepting an \accesstoken} 
\label{ssub:sniff_access_token_code}

In the \authcodeflow, a \textit{code} is returned by Google to the RP's Google sign-in endpoint (see step 6 in section \ref{subsubsec:authorization-code-flow}). No tokens are transmitted during the authorization procedure. After the RP receives the \textit{code}, it can use it to retrieve an \accesstoken from Google (steps 7/8 in \ref{subsubsec:authorization-code-flow}); it can then use the \accesstoken to retrieve user attributes from Google (step 9 in \ref{subsubsec:authorization-code-flow}). The RP then logs the user in to its website.

If an RP does not use SSL to protect communications with its Google sign-in endpoint, a passive web attacker may be able to intercept the \textit{code}. A passive web attacker cannot use the \textit{code} to retrieve an \accesstoken from Google, as it will not know the RP's \clientsecret (shared by the RP and Google). However, we observed that, of the RPs using the \authcodeflow, 6\% of their Google sign-in endpoints (i.e.\ 4 out of 69) return an \accesstoken to the user's browser instead of binding the user to the RP's session. As these RPs do not use SSL to protect the transfer of the \accesstoken, a passive web attacker is able to obtain the user's \accesstoken returned from the RP's Google sign-in endpoint.


\subsubsection{Stealing an \accesstoken via Cross-site Scripting} 
\label{ssub:steal_access_token_through_xss}

Google's ``automatic authorization granting'' feature \cite{DBLP:conf/ccs/SunB12} generates an authorization response automatically if the user has maintained a session with Google and has previously granted permission for the RP concerned.  Using this feature, an attacker might be able to steal a user \accesstoken by exploiting an XSS vulnerability in the RP or the browser.

To test the feasibility of such an attack, an exploit written in JavaScript was implemented. The exploit takes advantage of a recently revealed vulnerability in Android's built-in browser \cite{androidXSS14} which allows an attacker to conduct a universal XSS attack \cite{DBLP:conf/sac/KirdaKVJ06,DBLP:conf/ndss/NadjiSS09,DBLP:conf/ndss/VogtNJKKV07,DBLP:conf/icse/WassermannS08}. The exploit uses a browser \textbf{window.open} event to send a forged authorization request to Google's authorization server, within which \textit{response\_type=code} (see step 2 in \ref{subsubsec:authorization-code-flow}) is changed to \textit{response\_type=code token id\_token}. If the user is logged in to his or her Google account and has previously granted permission for this RP, Google automatically generates an authorization response without the involvement of the user; this response is appended as a URI fragment (\#) to the redirect URI (see step 5 in section \ref{subsubsec:authorization-code-flow}) and is sent back to the RP (see step 6 in section \ref{subsubsec:authorization-code-flow}). As the RP Google sign-in endpoint does not expect an URI fragment, a predefined error page will be generated by the RP (e.g.\ a `404 not found'  or `Failed connection' error). The exploiting JavaScript can now extract the authorization response from the URL of the error page and send it to its opener window, where the \textbf{window.open} event is triggered. The opener window then sends the \accesstoken to the attacker's server.

Unfortunately, our results show that all the RPs that adopt the \authcodeflow are vulnerable to this attack. The vulnerability affects all Android versions up to 4.4, which as of April 6, 2015 still accounted for 53.2\% of Android devices\footnote{\url{https://developer.android.com/about/dashboards/index.html?utm_source=suzunone}}.


\subsubsection{Privacy Issues} 
\label{ssub:pravicy_issues_code}

Unlike the \hybridserverflow, only a \textit{code} is submitted back to the RP's Google sign-in endpoint (see step 6 in section \ref{subsubsec:authorization-code-flow}). No user information (e.g.\ a Google ID or an \idtoken) is transmitted during the authorization procedure. However, user information leakage might nevertheless occur if the RP Google sign-in endpoint sends the user information directly to the user's browser without SSL protection.

Our study revealed that 16\% of RPs using the \authcodeflow (i.e.\ 11 out of 69) return user information to the browser directly without SSL protection. Thus a passive web attacker is able to intercept potentially sensitive user information, e.g.\ if the user is using an open Wi-Fi network (see section \ref{sec:adversary_model_and_related_work}).


\subsubsection{Session Swapping} 
\label{ssub:swession_swap_code}

If an RP using the \authcodeflow does not enable anti-CSRF measures (e.g.\ by appending a \textit{state} value bound to the browser session to the tokens) to protect its Google sign-in endpoint, a web attacker can launch a session swapping attack, precisely as described in \ref{ssub:swap_session} for the \hybridserverflow.

Unlike the session swapping attack in \ref{ssub:swap_session}, in the \authcodeflow only the GET method is used to submit the \textit{code} back to the RP's Google sign-in endpoint. This means that the attacker can simply insert the forged request in the \textbf{src} attribute of a \textbf{img} or \textbf{iframe} tag of an HTML document. When the victim user visits the malicious HTML, the browser will automatically send the request to the RP's Google sign-in endpoint using the GET method.

We found that 35\% of the RPs using the \authcodeflow (i.e.\ 24 out of 69) are vulnerable to this attack. However, as the \textit{code} is a one time value, the attacker has to update it every time the attack page is visited by a victim user. As a result such an attack is not as harmful as the session swapping attack in the \hybridserverflow, where an \accesstoken which can be used multiple times is submitted back to the RP's Google sign-in endpoint.


\subsubsection{Forcing a Login Using a CSRF attack} 
\label{ssub:forge_login_using_csrf}

A CSRF login attack operates in the context of an ongoing interaction between a target web browser (running on behalf of a target user) and a target RP\@. In such an attack, a malicious website somehow causes the target user's browser to initiate an OpenID Connect authorization request to the OP\@. Because of Google's ``automatic authorization granting'' feature, receiving such a request can cause the Google OP to generate an authorization response which is delivered to the RP without the involvement of the user. If the target user is currently logged in to Google, the browser will send cookies containing the target user's Google OP-generated tokens, along with the attacker-supplied authorization request, to the OP\@. The OP will process the malicious authorization request as if it was initiated by the target user, and will generate an authorization response and send it to the RP\@. The target browser could be made to send the spurious request in various ways; for example, a malicious site visited by the browser could use the HTML \textbf{img} tag's \textbf{src} attribute to specify the URL of a malicious request, causing the browser to silently use a GET method to send the request.

Our experiments have shown that 35\% of the RPs which adopt the \authcodeflow (i.e.\ 24 out of 69) are vulnerable to such an attack. One consequence of this attack is that an attacker can cause a victim user to log in to the RP, as long as the user has previously logged in to Google. This could damage the user experience of the RP website, as the victim user might dislike such a potentially annoying ``automatic login'' feature.


\section{Discussion} 
\label{sec:discussion}

OpenID Connect builds on top of the current web infrastructure, in which web application vulnerabilities (e.g.\ cross-site request forgeries, cross-site scripting) are common and have been widely exploited \cite{owasp_2013_top10}. The existence of these vulnerabilities exacerbates the threat posed by some of the implementation issues we have identified.

Most of the vulnerabilities described in section \ref{sec:studying_the_security_of_openid_connect} are caused by a combination of certain characteristics of the Google service and RP design decisions that appear to value simplicity over security. We next consider in greater detail how and why the various classes of vulnerability that we have identified
have arisen.

\subsection{Customising the Hybrid-Server-side Flow} 
\label{sub:customised_hybrid_server_flow}

According to the OpenID Connect specification \cite{openidconnect}, a \textit{code} must be returned by the OP to the RP's Google sign-in endpoint (see step 6 in section \ref{subsubsec:hybrid-server-side-flow}). However, as described in section \ref{sub:exploiting_the_hybrid_server_side_flow}, many RPs using the \hybridserverflow do not properly follow the specification, and in particular some RP Clients submit an \accesstoken, an \idtoken and/or a Google ID back to their Google sign-in endpoints. It appears that the OpenID Connect Specification only acts as a loose guideline for these RPs.

Meanwhile, the authorization request generated in the \hybridserverflow by RPs which use Google's OpenID Connect API will always include \textit{response\_type=code token id\_token}; therefore the authorization response which is sent to the RP Client contains a \textit{code}, \accesstoken and \idtoken. Unlike the \authcodeflow, where only a \textit{code} is returned to the RP's Google sign-in endpoint (see step 6 in section \ref{subsubsec:authorization-code-flow}) and no RP Client exists, this gives the RP the ability to customise their \hybridserverflow. In fact our experiments have shown that as many as 70\% of RPs (i.e.\ 23 out of 33) customise their \hybridserverflow. These customisations potentially improve the performance of the OpenID Connect system at the RP as well as enhancing the user experience, but they also risk introducing new vulnerabilities into the system, e.g.\ allowing an attacker to log in to the RP as any victim user (see section \ref{ssub:log_in_to_the_rp_as_the_victim_user}) and to impersonate the victim user using an \textit{access\_token} generated for another RP (see section \ref{ssub:impersonation_the_victim_user}). Moreover, as the \textit{code}, \accesstoken and \idtoken are returned by Google inside a HTML document, these values are also revealed to the user agent and hence to any applications (e.g.\ browser plug-ins), which might be able to access the user agent. If the plug-in or user agent has vulnerabilities which could allow an attacker to access these values, the attacker can steal the user's \accesstoken; for example a malicious plug-in which has the right to read the content of HTML pages could obtain the \accesstoken.


\subsection{Confusion over use of the \textit{state} value in the Hybrid-Server-side Flow} 
\label{sub:the_state_value_in_the_hybrid_server_flow}

When RP developers construct the authorization request using the Google OpenID Connect API, they only need to specify the RP \clientid and permission scope in their code, as the other values in the authorization request are handled by the API\@. These other values include the \textit{state}, which is used to bind the authorization response to the authorization request, thereby preventing CSRF attacks \cite{DBLP:conf/ccs/BarthJM08,jovanovic2006preventing,DBLP:conf/fc/MaoLM09,zeller2008cross}. This simplifies the job of the RP developers and makes support for Google's OpenID Connect easier to implement, but at the cost of increasing the attack surface and opening the protocol to new vulnerabilities. In order to understand how the API deals with the \textit{state} value, we implemented an RP using the Google OpenID Connect API\@. Surprisingly, we found that the \textit{state} value extracted by the RP Client is actually a null value; this means that Google itself fails to deliver the \textit{state} value to the RP Client, and hence the \textit{state} value cannot be used to mitigate the threat of a CSRF attack. We also observed that one of the RPs using the \authcodeflow sends a null \textit{state} value back to its Google sign-in endpoint.

As the \textit{state} value generated by the Google OpenID Connect API is not bound to the RP's session and cannot be extracted by the RP Client, another \textit{state} value which is bound to the session needs to be implemented to protect the RP's Google sign-in endpoint against a CSRF attack. However, 73\% of RPs using the \hybridserverflow fail to take this step. As a result they are all vulnerable to the session swapping attack described in section \ref{ssub:swap_session}.

Google does recommend RPs to use a \textit{state} value to protect their Google sign-in endpoint. However, examination of the Google OpenID Connect sample code \cite{googleopenidconnectserversideflow2015} reveals that Google has not included a \textit{state} value in its example of an RP Client-generated AJAX request, which is used to send data back to the RP \cite{connect_server_side_google15}.  The lack of a \textit{state} parameter in the sample code and the complexity of implementing anti-CSRF measures helps to explain why 73\% of the RPs using the \hybridserverflow are vulnerable to this attack.


\subsection{Automatic Authorization Granting} 
\label{sub:automatic_authorization_granting}
The ``automatic authorization granting'' feature of Google's implementation of OpenID Connect significantly enhances the user experience and system performance. Without this feature, the user would have to click the ``OK'' button in a popup window whenever he or she wished to log in to an RP, in order to grant authorization. However, this feature can also be harmful, since its use may allow an attacker to steal an \accesstoken (see section \ref{ssub:steal_access_token_through_xss}) and force a user log in to the RP (see section \ref{ssub:forge_login_using_csrf}).

We also observed that in the \hybridserverflow, iframes are used to manage the session \cite{OpenIDConnectSessionManagement2014} between the RP Client and the OP\@. Suppose that a user, who has previously both granted permission for the RP and logged in to his or her Google account, visits the RP login page which contains an iframe pointing to the authorization request. Because of the ``automatic authorization granting'' feature, the browser can use the GET method to retrieve the authorization response from Google without the involvement of the user. The user agent and any applications (e.g.\ plug-ins) which can access the user agent are able to extract the authorization response, which might expose the \hybridserverflow to new attacks.


\section{Recommendations} 
\label{sec:recommendations}

OpenID Connect has been deployed by many RPs and OPs, and it appears that increasing numbers of RPs supporting the Google service will implement OpenID Connect for SSO now that Google has shut down its OpenID service. However, our study has revealed serious vulnerabilities in existing systems, and there is a significant danger that these vulnerabilities will be replicated in future systems.

Below we  make a number of recommendations, directed at both RPs and OPs, designed to address the vulnerabilities we have identified. These recommendations primarily apply to RPs using the Google service and to the Google OP itself, but some may have broader applicability. There are two reasons for making these recommendations, namely both to try to address the problems that exist in current systems, and to help ensure that future systems are built in a more robust way.


\subsection{Recommendations for RPs} 
\label{sub:recommendations_for_rps}
When using the OpenID Connect system, especially in the case of the \hybridserverflow, the RP's developers are responsible for designing the RP Client action upon receiving an authorization response from the Google OP\@. As a result the security of the OpenID Connect system for the RP largely depends on the security expertise of the RP developers. We have the following recommendations for RPs.

\begin{itemize}
    \item \textbf{Do not customise the Hybrid Server-side Flow:} One of the reasons OpenID Connect is vulnerable to the attacks described in sections \ref{ssub:log_in_to_the_rp_as_the_victim_user} and \ref{ssub:impersonation_the_victim_user} is that the RPs customise the \hybridserverflow. In particular, instead of submitting a \textit{code} back to its Google sign-in endpoint, the RP Client running the UA submits an \accesstoken or Google ID, which is used by the RP to authenticate the user. Such a customised \hybridserverflow might improve the user experience and the efficiency of the RP website, but at the cost of exposing the system to new attacks. RPs must implement the OpenID Connect \hybridserverflow strictly conforming to the OpenID Connect Specification.
    \item \textbf{Deploy countermeasures against CSRF attacks:} One reason the OpenID Connect systems we have investigated are vulnerable to CSRF and session swapping attacks is that the RPs have not implemented any of the well-known countermeasures to such attacks. In order to prevent CSRF attacks, Google recommends RPs to include the \textit{state} parameter in the OpenID Connect authorization request and response, and RPs should follow this recommendation.
    \item \textbf{Do not use a constant or predictable \textit{state} value:} Some RPs include a fixed \textit{state} value in the OpenID Connect authorization request. If the \textit{state} value is fixed, it cannot be uniquely bound to the browser session, thereby allowing an attacker to successfully forge a response, since the RP cannot distinguish between a legitimate response produced by a valid user and a forged response produced by an attacker. Hence, in such a case, the inclusion of the \textit{state} value does not protect against CSRF attacks. Thus RPs must generate a non-guessable \textit{state} value which should be bound to the browser session so that the \textit{state} value can used to verify the validity of the response.
\end{itemize}

\subsection{Recommendations for OPs} 
\label{sub:recommendations_for_ops}

In an OpenID Connect SSO system, the OP designs the process and provides the API for RPs. An RP wishing to support a particular OP must therefore comply with the requirements of that OP, and so the OPs play a critical role in the system. We have the following recommendations for OPs (and in particular for Google).
\begin{itemize}
    \item \textbf{Remove the \textit{token} from the authorization request in the Hybrid Server Flow:} In the \hybridserverflow, the \textit{token} in the authorization request causes Google to return an \accesstoken to the RP Client. This allows RP Clients to submit an \accesstoken back to their Google sign-in endpoints, as was the case for 58\% of RPs using the \hybridserverflow that we investigated. This practice gives rise to a range of possible impersonation attacks. Sending the \accesstoken also creates further risks, since if the RP does not enable SSL to protect its Google sign-in endpoint, a passive network attacker could steal it. This would not only enable a malicious RP to impersonate a user to those RPs which submit an \accesstoken to the Google sign-in endpoint, but also allow the possibility of other misuses of this token, e.g.\ to compromise sensitive user data.
    \item \textbf{Add a \textit{state} value to the sample code:} OPs typically provide sample code to help RP developers make their website interact appropriately with the OP\@. As we discovered, Google does not include a \textit{state} value in its sample code for the \hybridserverflow. It seems reasonable to speculate that this is the main reason why 73\% of the RP-OP interactions we have analysed (see section \ref{ssub:swap_session}) are vulnerable to session swapping attacks. However, for cases where a \textit{state} value is included in Google's sample code, this number fell to 35\% (see section \ref{ssub:swession_swap_code}).
    \item \textbf{Allow the RP to specify the \textit{state} value in the Hybrid Server Flow:} The \textit{state} value in the authorization request of the  \hybridserverflow is automatically handled by the Google OpenID Connect API\@. However, the RP Client cannot extract the \textit{state} as it is a null value. As the \textit{state} value is not bound to the browser session, it does not protect the RP against CSRF attacks. It would probably be better to let the RP handle the \textit{state} value rather than the Google API handle it. In addition, Google should check the source code of its postmessage.js script to ensure that the \textit{state} value can be extracted by the RP Client.
\end{itemize}

\subsection{Notifying affected parties} 
\label{sub:notifying_affected_parties}

We reported the issues described in section \ref{ssub:log_in_to_the_rp_as_the_victim_user} to the three affected parties in February and also provided advice to help them fix the problem.  As of 30 July 2015, one of the three has fixed the problem, one has ignored our warning, and the third terminated support for the Google SSO service.  On April 17th 2015 we also notified Google of all the issues described in this paper. Google acknowledged the problem described in section \ref{sub:the_state_value_in_the_hybrid_server_flow} and notified their OpenID Connect group.

\section{Related Work} 
\label{sub:related_work}

OAuth 2.0, as the predecessor of OpenID Connect, has been analysed using formal methods. Pai et al.\ \cite{pai11} confirmed a security issue described in the OAuth 2.0 Thread Model \cite{oauth2threat} using the Alloy Framework \cite{alloy}. Chari et al.\ analysed OAuth 2.0 in the Universal Composability Security framework \cite{DBLP:journals/iacr/ChariJR11} and showed that OAuth 2.0 is secure if all the communications links are SSL-protected. Frostig and Slack \cite{frostig11} discovered a cross site request forgery attack in the Implicit Grant flow of OAuth 2.0, using the Murphi framework \cite{DBLP:conf/cav/Dill96}. Bansal et al.\ \cite{DBLP:journals/jcs/BansalBDM14} analysed the security of OAuth 2.0 using the WebSpi \cite{WebSpi} and ProVerif models \cite{ProVerif}. However, all this work is based on abstract models of OAuth 2.0, and so delicate implementation details are ignored.

Meanwhile, a range of work exploring the security properties of real-world implementations of OAuth 2.0 has also been conducted. Wang et al.\ \cite{DBLP:conf/sp/WangCW12} examined a number of deployed SSO systems, focussing on a logic flaw present in many such systems, including OpenID\@. In parallel, Sun and Beznosov \cite{DBLP:conf/ccs/SunB12} also studied deployed systems of OAuth 2.0 providing services in English. Li and Mitchell \cite{DBLP:conf/isw/LiM14} examined the security of deployed OAuth 2.0 systems providing services in Chinese. In parallel, Zhou and Evans \cite{DBLP:conf/uss/ZhouE14} conducted a large scale study of the security of Facebook's OAuth 2.0 implementation. Chen et al.\ \cite{DBLP:conf/ccs/ChenPCTKT14}, and Shehab and Mohsen \cite{DBLP:conf/codaspy/ShehabM14} have looked at the security of the implementation of OAuth 2.0 on mobile platforms. However, despite all the work on OAuth, very little research has been conducted on the security of OpenID Connect systems.

\section{Concluding Remarks} 
\label{sec:concluding_remarks}

In this paper, we have reported on the first field study of the security properties of Google's implementation of OpenID Connect. We examined the security of all 103 of the RPs that implement support for the Google OpenID Connect service from the GTMetrix list of the Top 1000 Sites. The methodology we used to discover vulnerabilities is similar to that used by Wang et al.\ \cite{DBLP:conf/sp/WangCW12} and Sun and Beznosov \cite{DBLP:conf/ccs/SunB12}, i.e.\ we analysed the HTTP messages transmitted between the RP and OP via the browser; however, our approach was different in two key respects. First, we added a further class of adversary to the threat model, in which a malicious RP tries to collect user \textit{access\_tokens} and then use them to impersonate the user to other RPs. Second, we focussed our study on OpenID Connect rather than OAuth 2.0 and other generic SSO systems. This has allowed us to to identify gaps between the implementation and specification of OpenID Connect, discover a number of vulnerabilities which allow an attack to log in to the RP as a victim user, and propose practical and useful improvements which can be adopted by all OpenID Connect RPs and OPs.

\bibliographystyle{plain.bst}
\bibliography{bibliography}

\end{document}